\documentclass[sigconf,nonacm]{acmart}

\AtBeginDocument{%
  }

\renewcommand\footnotetextcopyrightpermission[1]{}

\usepackage{multirow}
\usepackage{algorithm}
\usepackage{algpseudocode}

\usepackage{booktabs}

\begin{document}

\title{TRACE: Temporal Rule-Anchored Chain-of-Evidence on Knowledge Graphs for Interpretable Stock Movement Prediction}

\newcommand{\affmark}[1]{\textsuperscript{#1}}

\author{%
Qianggang Ding\affmark{1,2},
Haochen Shi\affmark{1,2},
Luis Castej\'on Lozano\affmark{3},
Miguel Conner\affmark{3},
Juan Abia\affmark{3},
Luis Gallego-Ledesma\affmark{3},
Joshua Fellowes\affmark{3},
Gerard Conangla Planes\affmark{3},
Adam Elwood\affmark{4},
Bang Liu\affmark{1,2,5}
\\[0.4em]
\affmark{1} DIRO \& Institut Courtois, Université de Montréal, Montreal, QC, Canada \\
\affmark{2} Mila -- Québec AI Institute, Montreal, QC, Canada \\
\affmark{3} Aily Labs, Madrid, Spain \\
\affmark{4} Aily Labs, Milan, Italy \\
\affmark{5} Canada CIFAR AI Chair
\\[0.4em]
{\small
\texttt{\{qianggang.ding, haochen.shi, bang.liu\}@umontreal.ca} \\
\texttt{\{luis.castejon, miguel.conner, juan.abia, luis.gallego, joshua.fellowes, gerard.conangla, adam.elwood\}@ailylabs.com}
}
}
\renewcommand{\shortauthors}{Ding et al.}


\begin{abstract}
We present a Temporal Rule-Anchored Chain-of-Evidence (TRACE) on knowledge graphs for interpretable stock movement prediction that unifies symbolic relational priors, dynamic graph exploration, and LLM-guided decision making in a single end-to-end pipeline. The approach performs rule-guided multi-hop exploration restricted to admissible relation sequences, grounds candidate reasoning chains in contemporaneous news, and aggregates fully grounded evidence into auditable \texttt{UP}/\texttt{DOWN} verdicts with human-readable paths connecting text and structure. On an S\&P~500 benchmark, the method achieves 55.1\% accuracy, 55.7\% precision, 71.5\% recall, and 60.8\% F1, surpassing strong baselines and improving recall and F1 over the best graph baseline under identical evaluation. The gains stem from (i) rule-guided exploration that focuses search on economically meaningful motifs rather than arbitrary walks, and (ii) text-grounded consolidation that selectively aggregates high-confidence, fully grounded hypotheses instead of uniformly pooling weak signals. Together, these choices yield higher sensitivity without sacrificing selectivity, delivering predictive lift with faithful, auditably interpretable explanations.
\end{abstract}



\maketitle

\section{Introduction}

Predicting stock price movements has long been a central problem in computational finance and data mining.  
Classical approaches~\cite{feng2018enhancing,long2019deep,yoo2021accurate,nelson2017stock,ding2024tradexpert} rely on time-series modeling of historical prices and macroeconomic indicators, while more recent methods~\cite{ding2020hierarchical,nguyen2015sentiment,xu2018stock} incorporate textual sentiment extracted from financial news and social media.  
Despite significant progress, existing models remain limited in two key dimensions.  
First, they typically treat heterogeneous information sources independently, aggregating features without modeling the relational dependencies among companies, events, and products.  
Second, large language model (LLM)-based approaches, while powerful in natural language understanding, often hallucinate unsupported facts and lack interpretability, posing risks in high-stakes financial domains.

Financial markets are inherently relational: the behavior of a single stock rarely depends only on its own fundamentals or sentiment, but rather on a complex web of interactions.  
Corporate actions such as mergers, acquisitions, partnerships, supply chain disruptions, management changes, or regulatory rulings propagate across industries and geographies, influencing not only the directly involved companies but also competitors, suppliers, and customers.  
For instance, an acquisition in the artificial intelligence sector may boost the acquiring company, benefit upstream hardware suppliers, and exert competitive pressure on rival firms.  
Similarly, a regulatory investigation into a pharmaceutical company can negatively affect its partners and contract manufacturers.  

Knowledge graphs have been extensively studied in Web research~\cite{huang2019knowledge,rossi2021knowledge,lin2018multi}, particularly in the domains of question answering, link prediction, and multi-hop reasoning. Existing financial forecasting models rarely integrate explicit graph reasoning, thereby missing an opportunity to connect unstructured text (e.g., news) with structured relations (e.g., investments, partnerships) in a unified framework.  
By bringing knowledge graph reasoning into financial prediction, we aim to bridge this gap and capture the inherently relational dynamics of markets.

In this paper, to the best of our knowledge, we introduce the \textbf{first temporal knowledge graph enhanced framework with LLMs for stock movement prediction}. Our approach is not a simple combination of existing reasoning paradigms, but a \emph{unified methodology} that integrates symbolic priors, dynamic graph exploration, and LLM-guided decision making into a coherent prediction pipeline.  

At the heart of our framework are the following two components: \textbf{(1) Rule-guided relational priors.}  
We extract relational motifs from historical knowledge graph snapshots that capture recurring financial interactions (e.g., investment activities anchored in news, partnerships linked to causal events).  
These mined patterns are not treated as standalone rules, but instead define the \emph{search space} and \emph{structural biases} for subsequent exploration.  
By shaping which relation sequences are admissible, they ensure that reasoning paths reflect financially meaningful dependencies.  
\textbf{(2) Graph-based reasoning with textual grounding.}  
Building on these priors, we design a guided exploration procedure that traverses the financial knowledge graph from each S\&P 500 stock at date $t$, following only temporally valid relations.  
Candidate reasoning paths connect stocks to events, counterparties, and ultimately to \texttt{TextSource} nodes such as news or filings.  
Each path is scored using a multi-factor function that combines rule coverage, temporal recency, structural diversity, and feedback from an LLM relation selector.  
This process yields evidence chains that are both semantically relevant and anchored in explicit textual evidence.  

The integration of these components produces more than the sum of its parts.  
Rule-guided priors reduce noise and prevent arbitrary exploration, while graph reasoning operationalizes these priors to discover diverse, news-grounded evidence chains.  
The final prediction stage aggregates path-level signals into binary UP/DOWN forecasts, each supported by explicit reasoning paths.  
This unified design enables \emph{both} predictive accuracy and interpretability, addressing the long-standing trade-off in financial forecasting between performance and transparency.

This work makes three main contributions:  

\begin{enumerate}
    \item We propose a unified framework named TRACE that embeds interpretable stock movement prediction within a financial temporal knowledge graph, integrating symbolic path mining, LLM-guided graph exploration, and LLM-validated decision making. 

    \item We design a rule-guided and text-grounded reasoning procedure where mined relational motifs constrain admissible paths and candidate reasoning chains are anchored in news articles, ensuring both economic relevance and temporal validity.  

    \item We construct a large-scale financial knowledge graph from 2022–2023 S\&P 500 data (42K nodes, 174K edges) and conduct extensive experiments on it, demonstrating that our framework outperforms strong baselines while providing interpretable UP/DOWN predictions supported by explicit reasoning paths.  

\end{enumerate}

\section{Related Work}
\label{sec:relateed}
\textbf{Graph-based Reasoning in Financial Forecasting.}
Traditional time-series or text-based financial forecasting models generally treat each stock independently, ignoring the web of inter-firm dependencies that propagate risk and information across sectors~\cite{yang2020finbert,araci2019finbert,ding2015deep,zhou2021informer,ariyo2014stock,bai2018empirical,fischer2018deep}. While graph-based modeling provides a remedy by explicitly capturing relational dependencies among entities such as companies, products, and events~\cite{yang2020finbert,araci2019finbert,ding2015deep,zhou2021informer,bai2018empirical,ariyo2014stock,fischer2018deep}. 
Early financial graph models, such as HATS~\cite{kim2019hats} and RSR~\cite{feng2019temporal}, introduced multi-relational graphs built from co-occurrence or textual correlations between firms and news. In contrast, later works, including MDGNN~\cite{qian2024mdgnn} and FEEKG~\cite{liu2024risk}, explicitly modeled temporal layers of events to predict stock trends.
On the other hand, inter-firm dependencies and macro events in financial markets evolve continuously, which naturally aligns with the temporal knowledge graph (TKG) based modeling~\cite{trivedi2017know,dasgupta2018hyte,jin2019recurrent}.
More recently, large language models have been employed to enhance the construction and reasoning capacity of financial graphs.  
For example, FinDKG~\cite{li2024findkg} leverages LLM-based extraction to build dynamic financial KGs, and FinGPT~\cite{wang2023fingpt} or BloombergGPT~\cite{wu2023bloomberggpt} provide strong language priors for financial text understanding.  However, these approaches largely rely on latent embedding fusion or text-based representation learning. 
In contrast, our framework performs explicit temporal reasoning on a financial knowledge graph.
\textbf{Rule Learning and Neuro-symbolic Reasoning.}
While graph-based modeling captures temporal dependences, it still heavily relies on implicit relational embeddings. These kinds of methods offer limited interpretability into how relations drive predictions. To enhance structural interpretability, another line of work focuses on explicit rule learning and neuro-symbolic reasoning. Traditional inductive rule learning framework such as AMIE~\cite{galarraga2015fast}, AnyBURL~\cite{meilicke2020reinforced} focuses on mining frequent relational patterns from knowledge graphs to provide transparent logical rules for inference. However, differentiable methodologies integrate rule learning with gradient-based optimization, bridging symbolic and neural paradigms~\cite{yang2017differentiable,qu2020rnnlogic}. More recent work explores hybrid systems by marrying rule induction with pre-trained language models to enhance reasoning over KGs~\cite{saeed2021rulebert}. Different from these works, our framework treats mined relational rules as dynamic search priors within a temporal financial knowledge graph. Instead of applying static rules or fusing embeddings, we employ rules to constrain admissible reasoning paths under as-of temporal constraints, ensuring causal validity. 

\textbf{LLM-KG Hybrid Reasoning.} 
Building upon advances in rule-based and neuro-symbolic reasoning, a growing line of work explores combining LLMs with structured KGs to enhance factual grounding and reasoning interpretability~\cite{jiang2023structgpt,wang2023boosting,baek2023knowledge,xie2022unifiedskg,sun2023think,ma2024think,luo2024graph}.  For instance, Think-on-Graph (ToG)~\cite{sun2023think, ma2024think} deploys LLMs as reasoning agents that iteratively traverse a KG through beam search~\cite{Jurafsky2009-ei,10.1145/3539618.3591698,sun2023beamsearchqa,liumakes} and construct explicit relational chains to reduce hallucination. Graph-Constrained Reasoning (GCR)~\cite{luo2024graph} further improves structural faithfulness by constraining LLM decoding via a KG-guided trie.  Another line of work extends this paradigm by integrating LLM reasoning with KG-derived knowledge adaptation. For example, Chain-of-Knowledge (CoK)~\cite{zhang2024chain} aligns LLM reasoning with KG paths by training the model to reason via structured paths on the dataset of mined rules. GenTKG~\cite{liao2023gentkg} extends this idea to temporal domains, coupling retrieval-augmented generation with time-aware relational reasoning to enable generative forecasting over evolving KGs.  Based on this idea, RAG-KG-IL~\cite{yu2025rag} integrates retrieval-augmented generation with incremental KG learning in a multi-agent framework to facilitate continual knowledge updating and adaptive reasoning across heterogeneous sources. 
In contrast, our framework unifies symbolic priors, dynamic graph exploration, and LLM-guided decision making on a temporal financial KG.

\begin{figure}[t]
\centering
\includegraphics[width=0.48\textwidth]{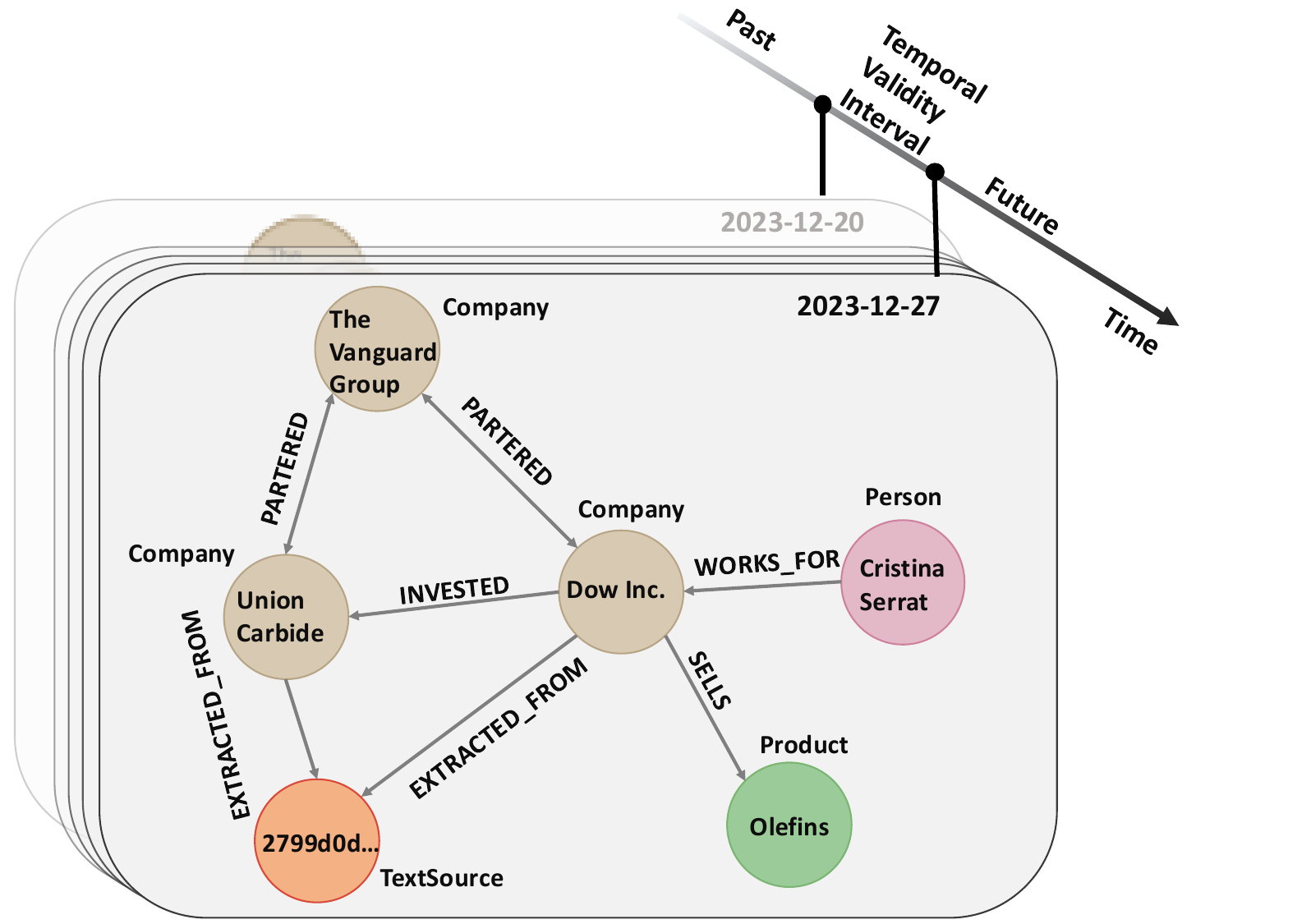}
\caption{Structure of the temporal financial knowledge graph showing entity types, relation categories, and temporal validity intervals.}
\label{fig:kg_structure}
\end{figure}

\section{Methodology}

\subsection{Problem Formulation}
\label{sec:problem_formulation}

We study the task of \emph{stock movement prediction} using dynamic knowledge graphs (KGs). 
Let $\mathcal{G}_t = (\mathcal{E}_t, \mathcal{R}_t, \mathcal{F}_t)$ denote a temporal knowledge graph snapshot at trading day $t$, where:
\begin{itemize}
    \item $\mathcal{E}_t$ is the set of entities, including \texttt{TextSource}, \texttt{Event}, \texttt{Product}, \texttt{Financial}, \texttt{Company}, and \texttt{Person} nodes;
    \item $\mathcal{R}_t$ is the set of relation types, including \texttt{EXTRACTED\_FROM}, \texttt{SELLS}, \texttt{INVESTED\_IN}, \texttt{PARTNERED}, \texttt{SUED}, etc;
    \item $\mathcal{F}_t \subseteq \mathcal{E}_t \times \mathcal{R}_t \times \mathcal{E}_t$ is the set of temporal triples $(e_h, r, e_t)$ valid at time $t$.
\end{itemize}

Each stock $s \in \mathcal{S}$ (where $\mathcal{S}$ is the set of S\&P 500 stocks) is represented by a \texttt{Company} node and linked to its ticker symbol through a mapping function 
\[
    \pi: \mathcal{E}_t \to \mathcal{T}, \quad \pi(s) = \text{ticker of stock $s$}.
\]

\paragraph{\textbf{Text sources and as-of constraint.}}  
We denote by $\mathcal{X}$ the set of \texttt{TextSource} nodes (financial news, filings, reports).  
Each $x \in \mathcal{X}$ has an associated publication date $\tau(x)$.  
To avoid future leakage, only edges and facts with timestamps $\leq t$ are accessible when constructing $\mathcal{G}_t$.  
Formally, a triple $(e_h, r, e_t) \in \mathcal{F}_t$ is valid iff
\[
    \texttt{valid\_from}(e_h, r, e_t) \leq t \ \ \wedge \ \ 
    \big( \texttt{valid\_to}(e_h, r, e_t) \geq t \ \text{or} \ \texttt{valid\_to} = \varnothing \big).
\]

\paragraph{\textbf{Rules.}}  
We assume a pre-mined rule bank $\mathcal{R}^\ast = \{ \rho_1, \ldots, \rho_M \}$ obtained via automated rule mining methodology.
Each rule $\rho$ is defined with both a body pattern and a predictive head:
\[
    \rho: \underbrace{r_1(X, Z_1) \wedge r_2(Z_1, Z_2) \wedge \cdots \wedge r_L(Z_{L-1}, Y)}_{\text{body}} \Rightarrow \underbrace{\text{predict\_direction}(X, \ell)}_{\text{head}},
\]
where each $r_i \in \mathcal{R}_t$ and $\ell \in \{\text{UP}, \text{DOWN}\}$.
These rules act as \emph{label-predictive priors} for efficient path exploration and scoring, with explicit confidence values from historical validation.

\paragraph{\textbf{Prediction task.}}  
For each trading day $t$ within the evaluation window $[t_{\text{start}}, t_{\text{end}}]$, and for each stock $s \in \mathcal{S}$, the model predicts whether the stock will go 
\emph{up} or \emph{down} in the next trading day.

Let $p_{s,t,1}$ denote the forward return of stock $s$ from day $t$ to day $t+1$ (next trading day), computed from adjusted close prices:
\[
    p_{s,t,1} = \frac{\text{adj\_close}_{s,t+1} - \text{adj\_close}_{s,t}}{\text{adj\_close}_{s,t}}.
\]

The ground-truth label is then defined as:
\[
    y_{s,t,1} =
    \begin{cases}
        \text{UP}, & \text{if } p_{s,t,1} > 0, \\
        \text{DOWN}, & \text{if } p_{s,t,1} \leq 0.
    \end{cases}
\]

\paragraph{\textbf{Objective.}}  
The goal of our knowledge graph reasoning framework is to learn a reasoning function 
\[
    f: (\mathcal{G}_t, s, \mathcal{R}^\ast, h) \mapsto \hat{y}_{s,t,h},
\]
that outputs a prediction $\hat{y}_{s,t,h} \in \{\text{UP}, \text{DOWN}\}$ along with a set of interpretable reasoning paths $\mathcal{P}_{s,t,h}$ extracted from $\mathcal{G}_t$.  

\begin{figure*}[t]
\centering
\includegraphics[width=1.02\textwidth]{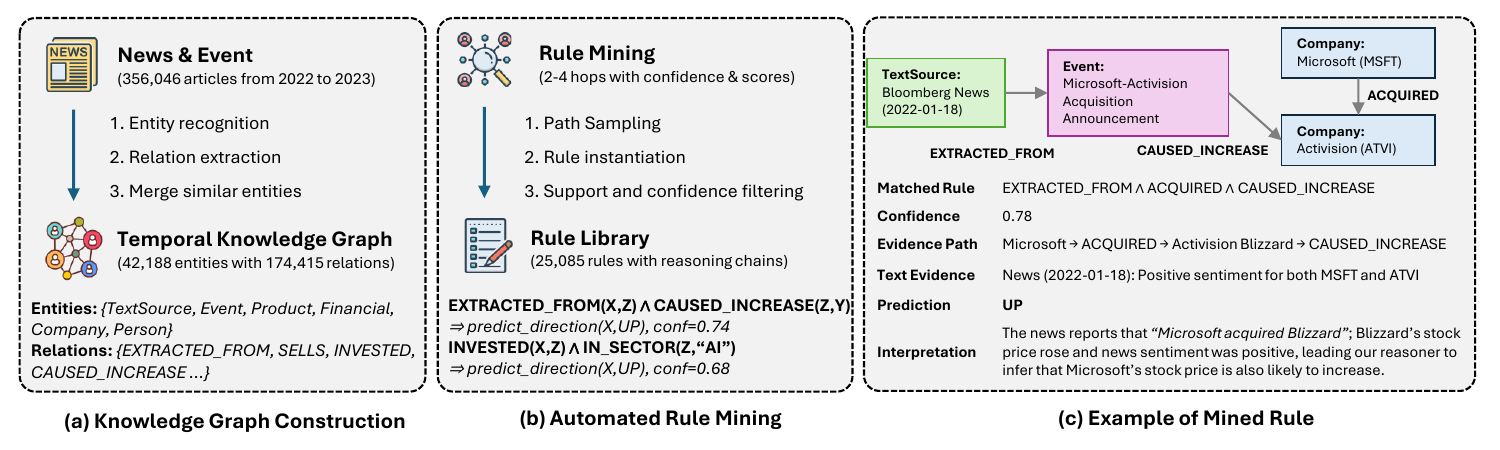}
\caption{Overview of our knowledge graph reasoning framework showing the integration of automated rule mining and temporal graph reasoning.}
\label{fig:framework}
\end{figure*}

\subsection{Knowledge Graph Construction}
\label{sec:kg_construction}

Our methodology relies on a dynamic financial knowledge graph (see Figure~\ref{fig:kg_structure}) that integrates heterogeneous data sources into a unified temporal graph.  
The KG is designed to capture both structured corporate relations and unstructured textual signals (e.g., news, filings), while preserving temporal validity.

\paragraph{\textbf{Entity types.}}  
The KG contains multiple categories of entities, each uniquely identified by a \texttt{uid}:
\begin{itemize}
    \item \textbf{TextSource} ($10{,}685$ nodes): financial news articles, earnings reports, and filings, each associated with a publication timestamp.
    \item \textbf{Event} ($9{,}174$ nodes): discrete corporate or market events such as mergers, lawsuits, or regulatory actions.
    \item \textbf{Product} ($8{,}973$ nodes): company products, technologies, and services.
    \item \textbf{Financial} ($5{,}991$ nodes): financial instruments, indicators, or assets other than S\&P 500 stocks.
    \item \textbf{Company} ($4{,}865$ nodes): corporate entities, including S\&P 500 constituents and their counterparties.
    \item \textbf{Person} ($2{,}500$ nodes): key executives, founders, and individuals relevant to market events.
\end{itemize}

\paragraph{\textbf{Relations.}}  
Edges represent typed relations between entities.  
Each relation is stored as a temporal triple $(e_h, r, e_t, t_{\text{from}}, t_{\text{to}})$, where $e_h, e_t \in \mathcal{E}$, $r \in \mathcal{R}$, and $[t_{\text{from}}, t_{\text{to}}]$ is the validity interval.  
The relation set includes:
\begin{itemize}
    \item \texttt{EXTRACTED\_FROM} ($125{,}276$ edges): anchors entities (e.g., Companies, Events, Products) to their originating \texttt{TextSource}, thereby grounding structured nodes in textual evidence.
    \item \texttt{SELLS} ($13{,}003$) and \texttt{INVESTED\_IN} ($11{,}706$): capture product sales relations and investment activities.
    \item \texttt{CAUSED\_INCREASE} ($8{,}250$) and \texttt{CAUSED\_DECLINE} ($3{,}891$): encode causal event-to-financial effects extracted from text.
    \item \texttt{PARTNERED} ($3{,}933$), \texttt{DIVESTED} ($3{,}694$), and \texttt{ACQUIRED} ($1{,}004$): represent corporate restructuring and partnerships.
    \item \texttt{WORKS\_FOR} ($3{,}012$): links \texttt{Person} nodes to their affiliated companies.
    \item \texttt{LICENSED} ($415$), \texttt{SETTLED} ($153$), and \texttt{SUED} ($78$): describe legal, contractual, and licensing relations.
\end{itemize}

\paragraph{\textbf{Anchoring to text.}}  
Each \texttt{TextSource} node $x$ has metadata including title, publisher, URL, and publication date $\tau(x)$.  
Entities mentioned in $x$ are connected via \texttt{EXTRACTED\_FROM} edges, allowing reasoning paths to be grounded in explicit textual evidence.  

\paragraph{\textbf{Rule integration.}}  
A rule bank $\mathcal{R}^\ast$ mined via automated rule extraction is integrated into the KG reasoning process.  
While rules are not embedded directly in $\mathcal{G}_t$, their relational patterns constrain search trajectories and guide scoring during exploration.


\subsection{Automated Rule Mining}
\label{sec:rule_mining}

To guide efficient reasoning in our framework, we integrate a rule bank mined from the financial knowledge graph using automated pattern extraction methodology (see Figure~\ref{fig:framework}).
The purpose of rule mining is to extract recurring relational patterns that serve as \emph{priors} for our knowledge graph reasoning framework, improving both efficiency and interpretability.

\paragraph{\textbf{Rule definition.}}
A rule is defined as a sequence of relational atoms (body) connecting entities, with an explicit head that maps to stock movement prediction. Please refer to \ref{sec:problem_formulation} for rule definition.

\paragraph{\textbf{Mining procedure.}}
Following an automated mining framework, we employ a breadth-first enumeration of paths in the KG to construct candidate rules:
\begin{enumerate}
    \item \textbf{Path sampling.} From all triples $(e_h, r, e_t)$ in the KG, we construct 2-hop paths, which are recursively combined into 3-hop and 4-hop paths.
    \item \textbf{Rule instantiation.} Each path pattern is generalized into a rule body by replacing entities with variables. For each body pattern, we create both UP and DOWN head variants.
    \item \textbf{Support and confidence filtering.}
    For a rule candidate $\rho$ with head $\text{predict\_direction}(X, \ell)$, let $x$ denote the number of stock-day instances where the body pattern is satisfied, and let $y$ denote the number of such instances where the actual stock movement matches the predicted direction $\ell$.
    The confidence is defined as:
    \[
        \text{conf}(\rho) = \frac{y}{x}.
    \]
    Rules with support below a frequency threshold or with $\text{conf}(\rho) < \tau_{\text{mine}}$ are discarded, where $\tau_{\text{mine}} = 0.60$ is our mining threshold.
    \item \textbf{Pruning.} Rules that are redundant (subsumed by shorter rules) or overly specific (support too small) are removed to avoid sparsity.
\end{enumerate}

\paragraph{\textbf{Examples.}}  
Illustrative mined rules include:
\begin{align*}
    &\texttt{INVESTED\_IN}(X,Z) \wedge \texttt{IN\_SECTOR}(Z,\text{``AI''}) \\
    &\qquad \Rightarrow \quad \text{signal for positive movement of } X, \\
\end{align*}

Automated rule mining transforms raw KG structure into a compact library of relational patterns that guide our knowledge graph reasoning framework.
This synergy between symbolic rules and neural reasoning reduces search space, mitigates hallucination, and grounds predictions in financially meaningful causal chains.

\begin{figure*}[t]
\centering
\includegraphics[width=0.95\textwidth]{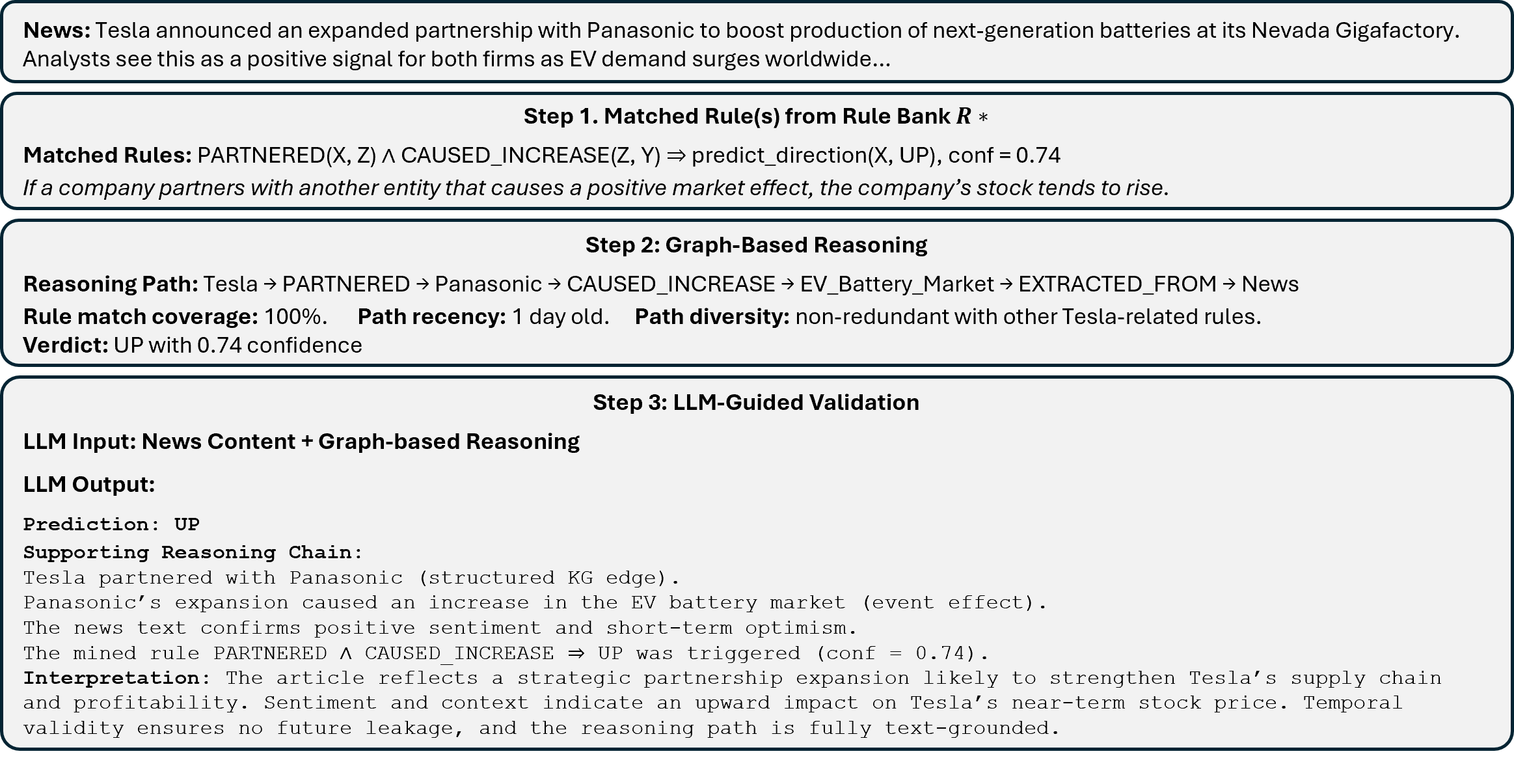}
\caption{Example of interpretable reasoning paths showing how our knowledge graph reasoning framework connects stocks to textual evidence.}
\label{fig:interpretability_examples}
\end{figure*}

\subsection{Graph-Based Exploration}
\label{sec:tog}

Our framework centers on graph-based exploration, which is an iterative, rule- and LLM-guided traversals of the financial KG that surface interpretable reasoning paths for stock movement prediction. An example of interpretable reasoning paths is shown in Figure \ref{fig:interpretability_examples} and the algorithm framework is illustrated in Alg.~\ref{alg:rule_guided_tkg}.

\paragraph{\textbf{Search initialization.}}  
For each trading day $t \in [t_{\text{start}}, t_{\text{end}}]$ and each stock $s \in \mathcal{S}$, we initialize exploration at the corresponding \texttt{TextSource} node.  
The \emph{as-of constraint} enforces that only edges and \texttt{TextSource} nodes with timestamps $\leq t$ are accessible.  
Thus, every reasoning path $\mathcal{P}_{s,t}$ is grounded in information available at the prediction time.

\paragraph{\textbf{Beam search.}}  
We employ beam search to explore the KG while controlling combinatorial explosion.  
At depth $d$, each partial path $P = (e_0 \xrightarrow{r_1} e_1 \xrightarrow{r_2} \cdots \xrightarrow{r_d} e_d)$ can be extended by candidate relations $\{r_{d+1}\}$ leading to neighbors of $e_d$.  
The beam maintains the top-$K$ paths ranked by a scoring function (see below).  
Search proceeds until either (i) a \texttt{TextSource} node is reached, or (ii) the maximum depth $D_{\max}$ is reached.

\paragraph{\textbf{Rule-Grounded Path Expansion.}}
A key innovation of our methodology is the integration of rule-grounded reasoning during graph exploration. For each partial path $P$ during expansion, we attempt to match it against the body of known label-predictive rules in $\mathcal{R}^\ast$:

\begin{enumerate}
    \item \textbf{Rule Matching:} Given current partial path $P = (e_0 \xrightarrow{r_1} e_1 \xrightarrow{r_2} \cdots \xrightarrow{r_d} e_d)$, we find all rules $\rho \in \mathcal{R}^\ast$ whose body sequences match the relation pattern $\{r_1, r_2, \ldots, r_d\}$.

    \item \textbf{Hypothesis Generation:} When a rule's body is fully grounded and the rule confidence exceeds our hypothesis threshold ($\text{conf}(\rho) > \tau_{\text{hyp}}$ where $\tau_{\text{hyp}} = 0.60$), we treat this as a label-predictive hypothesis. The hypothesis consists of:
    \begin{itemize}
        \item The grounded rule $\rho$ with confidence $\text{conf}(\rho) > \tau_{\text{hyp}}$ and predicted direction $\ell$
        \item The complete evidence path $P$ supporting the rule body
        \item Associated TextSource nodes providing textual evidence
    \end{itemize}
\end{enumerate}

This rule-grounded approach ensures that exploration is not merely traversing the graph structure, but actively seeking evidence that validates high-confidence predictive patterns discovered during rule mining.

\paragraph{\textbf{Relation selection.}}
Given the current path prefix $\{r_1, \ldots, r_d\}$ and the set of candidate relations from $e_d$, we apply a two-stage filter:
\begin{enumerate}
    \item \textbf{Rule-based admissibility:} the relation sequence must be a prefix of at least one rule body $\rho \in \mathcal{R}^\ast$;
    \item \textbf{LLM-based pruning:} if enabled, an LLM relation selector evaluates candidate relations in context (node type, as-of date, neighbor statistics) and returns a subset with scores $\{0,1,2\}$ indicating relevance.
\end{enumerate}
If the LLM selector is disabled, a heuristic scorer based on neighbor frequency and recency is used.

\paragraph{\textbf{Path scoring.}}
We use \emph{per-depth lexicographic} ranking. A path $P$ is admissible iff (i) its relation sequence is a prefix of some rule in $\mathcal{R}^\ast$, (ii) all edges are valid as of $t$, and (iii) $\mathrm{len}(P)\le D_{\max}$.

At depth $d$, compute four signals for each admissible $P$, each scaled to its \emph{percentile} among candidates at depth $d$:
$\mathrm{hyp}_\tau(P)\!\in\!\{0,1\}$ (completes a mined rule with confidence $\ge\tau$),
$\mathrm{cov}(P)\!\in\![0,1]$ (best rule-body coverage),
$\mathrm{rec}(P)\!\in\![0,1]$ (freshest timestamp on $P$),
$\mathrm{ahub}(P)\!\in\![0,1]$ (penalizes high-degree nodes; higher is less hub-biased).

Rank by the lexicographic key
\[
K(P)=\big(-\mathrm{hyp}_\tau(P),\; -\mathrm{cov}(P),\; -\mathrm{rec}(P),\; -\mathrm{ahub}(P),\; \mathrm{len}(P),\; \mathrm{hash}(P)\big),
\]
ascending, where minus signs mean “higher is better” and $\mathrm{hash}(P)$ is a deterministic node-sequence hash for stable tie-breaking. This scheme prioritizes completion of high-confidence rules, then favors paths that best align with rule bodies, are temporally recent, and avoid hubs.

\paragraph{\textbf{Early stopping and hypothesis validation.}}
Search halts early when a path reaches a \texttt{TextSource} node and forms a complete label-predictive hypothesis. This occurs when:
\begin{itemize}
    \item A rule's body is fully grounded in the current graph snapshot
    \item The rule has high confidence ($\text{conf}(\rho) > 0.6$)
    \item TextSource evidence supports the complete reasoning chain
\end{itemize}
When such a hypothesis is formed, it is added to the set of candidate explanations for the stock prediction. If an extension lacks supporting facts or cannot complete any high-confidence rule, the trial-and-error mechanism discards the branch and backtracks to alternative candidates.

\subsection{Verdict Prediction}
\label{sec:verdict}

After graph exploration, we obtain for each stock $s$ on day $t$ a set of candidate reasoning paths $\mathcal{P}_{s,t,1}$ for our primary 1-day prediction horizon (with support for other horizons $\mathcal{P}_{s,t,h}$ where $h \in \{3,5,10,30\}$).
Each path terminates at one or more \texttt{TextSource} nodes and is aligned with a subset of rules from the mined rule bank $\mathcal{R}^\ast$.
The objective of this stage is to consolidate these candidate paths into a binary verdict $\hat{y}_{s,t,1} \in \{\text{UP}, \text{DOWN}\}$ together with a confidence score for our primary focus.

\paragraph{\textbf{Primary verdict generation.}}
The final prediction is determined by aggregating confidence scores from high-confidence rule-based hypotheses $\mathcal{H}_{s,t,h} = \{(P_i, \rho_i, \text{conf}(\rho_i))\}$ generated during graph exploration. Each hypothesis consists of a grounded rule $\rho_i$ with confidence $\text{conf}(\rho_i) > \tau_{\text{hyp}}$ and predicted direction $\ell_i$. We aggregate as follows:
\begin{align*}
    \text{conf}_{\text{UP}}(s,t,h) &= \max_{\substack{(P_i, \rho_i) \in \mathcal{H}_{s,t,h} \\ \ell_i = \text{UP}}} \text{conf}(\rho_i), \\
    \text{conf}_{\text{DOWN}}(s,t,h) &= \max_{\substack{(P_i, \rho_i) \in \mathcal{H}_{s,t,h} \\ \ell_i = \text{DOWN}}} \text{conf}(\rho_i).
\end{align*}
The final prediction is chosen as the label with higher aggregated confidence:
\[
    \hat{y}_{s,t,h} =
    \begin{cases}
        \text{UP}, & \text{if } \text{conf}_{\text{UP}} \geq \text{conf}_{\text{DOWN}}, \\
        \text{DOWN}, & \text{otherwise}.
    \end{cases}
\]

\paragraph{\textbf{Path evaluation.}}
Each candidate path $P \in \mathcal{P}_{s,t,1}$ is examined by a large language model (LLM) under strict prompting constraints, producing a binary label and confidence $(y_P, p_P)$. This evaluation serves for explanation ranking and interpretability scoring.

\paragraph{\textbf{Interpretability of verdicts.}}  
Each final decision is accompanied by the top-$M$ supporting reasoning paths that contributed to the outcome.  
For instance, a verdict of ``UP'' may be supported by:
\[
    \texttt{Company A--INVESTED\_IN-->Startup Z--EXTRACTED\_FROM-->News Article},
\]
which aligns with the rule pattern $\rho: \ \texttt{INVESTED\_IN}(X,Z) \quad \wedge \quad $ $\texttt{EXTRACTED\_FROM}(Z,T)$.  
This explicit linkage between the verdict, the underlying rule, and the textual anchor enables transparent, human-auditable financial forecasts.

\section{Experiments}
\label{sec:experiments}

We conduct comprehensive experiments to evaluate our temporal knowledge graph reasoning approach on real-world financial prediction tasks. Our evaluation demonstrates both superior predictive performance and high-quality interpretable explanations compared to state-of-the-art baselines.

\subsection{Experimental Setup and Datasets}
\label{sec:datasets}

\paragraph{\textbf{Financial Knowledge Graph Construction.}}
We construct a temporal financial knowledge graph for S\&P 500 companies (2022 Jan to 2023 Jan), containing 42,188 entities and 174,415 timestamped edges. Refer to Sec.~\ref{sec:kg_construction}. We use daily OHLCV data for all 487 S\&P 500 stocks (187,452 observations), with split/dividend adjustments and outlier filtering. Binary labels are defined from forward returns: $y_{s,t,1} = \mathbb{I}[P_{s,t+1} > P_{s,t}]$, yielding balanced UP/DOWN distributions (51.2\%/48.8\%). We enforce strict as-of-date constraints and use walk-forward validation with daily predictions.

\subsection{Evaluation Metrics}
\label{sec:metrics}

We evaluate using standard classification metrics: Accuracy, Precision, Recall, and F1-Score. For interpretability, we measure Path Coverage, Rule Utilization, Evidence Grounding, and Path Diversity.

\subsection{Baseline Methods}
\label{sec:baselines}
We evaluate representative baselines under the same as-of-date protocol:
\begin{itemize}
    \item \textbf{Historical Momentum}: Predicts next-day direction from recent price trend using a standard momentum signal. The lookback window is selected on validation and applied out-of-sample.
    \item \textbf{XGBoost}~\cite{vuong2022stock}: Gradient-boosted trees trained on daily OHLCV features with lightweight technical indicators. Class weighting and all hyperparameters are tuned on validation without look-ahead.
    \item \textbf{News Sentiment (Qwen QwQ-32B)}~\cite{yang2025qwen3,nguyen2015sentiment}: A news-only model that uses Qwen to score per-stock daily sentiment from contemporaneous articles; scores are aggregated per day and thresholded into UP/DOWN.
    \item \textbf{T-GNN}~\cite{xiang2022temporal}: A temporal graph neural network over our financial knowledge graph, performing message passing across time under identical splits and temporal constraints.
    \item \textbf{Agent frameworks}: \emph{Mars}~\cite{chen2025mars}, \emph{StockAgent}~\cite{zhang2024ai}, and \emph{FinMem}~\cite{yu2025finmem} are self-implemented to conform to our classification harness. Their outputs are standardized rule-based proxies for agent-style pipelines and are included for completeness; the reported numbers should not be interpreted as the original systems' official results.
    \item \textbf{Ours}: The proposed temporal KG reasoning model with rule-guided multi-hop exploration and path aggregation.
\end{itemize}
All methods use identical data splits, features restricted by information dates, and the same evaluation protocol. For backtesting evaluation, we construct Top-10 buy-and-hold baskets per method (equal-weight, no rebalancing) and report an equal-weight all-stocks benchmark. Unless noted otherwise, the training window is 2022-01-01 to 2023-01-01 and the evaluation window is 2023-01-01 to 2024-01-01. All LLM experiments use the Qwen QwQ-32B model~\cite{yang2025qwen3}.

\subsection{Main Experimental Results}
\label{sec:main_results}

Table~\ref{tab:main_results} presents performance comparison across baseline methods for 1-day horizon stock movement prediction.

\begin{table}[t]
\centering
\caption{Performance comparison for 1-day stock movement prediction. Best results in \textbf{bold}. }
\label{tab:main_results}
\begin{tabular}{l|cccc}
\toprule
Method & Accuracy & Precision & Recall & F1-Score \\
\midrule
Historical Momentum      & 0.517 & 0.524 & 0.709 & 0.603 \\
XGBoost                  & 0.529 & 0.535 & 0.678 & 0.598 \\
News Sentiment (Qwen)    & 0.524 & 0.529 & 0.636 & 0.594 \\
T-GNN                    & 0.528 & 0.534 & 0.685 & 0.600 \\
\midrule
Mars       & 0.539 & 0.544 & 0.664 & 0.598 \\
StockAgent  & 0.539 & 0.545 & 0.648 & 0.592 \\
FinMem     & 0.544 & 0.550 & 0.647 & 0.594 \\
\midrule
\textbf{TRACE (Ours)}            & \textbf{0.551} & \textbf{0.557} & \textbf{0.715} & \textbf{0.608} \\
\bottomrule
\end{tabular}
\end{table}

\begin{table}[t]
\centering
\caption{Detailed ablation study showing the contribution of individual components to overall performance.}
\label{tab:detailed_ablation}
\begin{tabular}{l|ccc}
\toprule
\multirow{2}{*}{Configuration} & \multicolumn{3}{c}{Performance Metrics} \\
\cmidrule(lr){2-4}
 & Accuracy & Precision & Recall \\
\midrule
Full Model & \textbf{0.551} & \textbf{0.557} & \textbf{0.715}  \\
\midrule
- Temporal Constraints & 0.539 & 0.544 & 0.534 \\
- Rule Mining & 0.542 & 0.547 & 0.537 \\
- Multi-hop Reasoning & 0.532 & 0.537 & 0.527  \\
- Path Aggregation & 0.548 & 0.553 & 0.543 \\
- LLM Relation Selection & 0.545 & 0.550 & 0.540\\
\midrule
Random Classifier & 0.502 & 0.501 & 0.500\\
\bottomrule
\end{tabular}
\end{table}

\begin{table*}[t]
\centering
\caption{Top-10 buy\&hold portfolio metrics (2023-01-01 to 2024-01-01). Each method forms a long-only, equal-weight basket of its Top-10 tickers without rebalancing. Best values in \textbf{bold}.}
\label{tab:backtest_metrics}
\begin{tabular}{l|ccccc}
\toprule
Method & Total Return & Sharpe & Ann. Vol & Max DD & Win Rate \\
\midrule
Equal-Weight All & 16.7\% & 1.17 & 14.1\% & -12.9\% & 52.8\% \\
Historical Momentum & 8.7\% & 0.62 & 15.4\% & -17.3\% & 54.4\% \\
XGBoost & 27.0\% & 1.31 & 19.9\% & -14.8\% & 53.2\% \\
News Sentiment (Qwen) & 9.2\% & 0.67 & 14.8\% & -12.2\% & 52.8\% \\
T-GNN & 3.5\% & 0.33 & \textbf{13.3\%} & -12.5\% & 51.2\% \\
Mars & 14.2\% & 0.98 & 14.7\% & -17.3\% & 54.8\% \\
StockAgent & 26.4\% & 1.43 & 17.7\% & -15.0\% & 52.4\% \\
FinMem & 13.1\% & 0.88 & 15.5\% & -15.7\% & 55.2\% \\
\textbf{Ours} & \textbf{41.7\%} & \textbf{2.00} & 18.5\% & \textbf{-11.7\%} & \textbf{56.0\%} \\
\bottomrule
\end{tabular}
\end{table*}

\paragraph{\textbf{Predictive Performance Analysis.}}
Table~\ref{tab:main_results} shows that our method attains the best one day performance with 55.1\% accuracy, 55.7\% precision, 71.5\% recall, and 60.8\% F1. This corresponds to gains of 2.2 accuracy points over XGBoost at 52.9\% and 0.7 points over FinMem at 54.4\%. The improvement aligns with the methodology. First, rule-guided exploration restricts each expansion to prefixes of mined rule bodies and promotes hypothesis formation only when a high confidence rule is fully grounded with supporting TextSource evidence, which concentrates search on economically meaningful chains rather than arbitrary multi hop traversals. Second, the per depth lexicographic ranking prioritizes completion of high confidence rules, then body coverage, penalizes hubs, and uses stable tie breaking, which elevates paths that instantiate strong priors over high degree shortcuts. Third, the verdict stage aggregates high confidence grounded hypotheses into a single label and confidence, providing a selective fusion of heterogeneous evidence rather than uniform pooling. Together, these components yield higher sensitivity without a loss in selectivity, reflected by a recall gain of 3.0 points and an F1 gain of 0.8 points over T-GNN, alongside the overall accuracy improvement under the same evaluation protocol.

\section{Analysis}
\label{sec:analysis}

This section demonstrates experiments and results for our framework's interpretability quality, human expert analysis, ablation studies, backtesting analysis, and counterfactual analysis.

\subsection{Interpretability Quality Analysis}
\label{sec:interpretability}

A key advantage of our approach is the generation of interpretable reasoning paths that support each prediction.

\paragraph{\textbf{Path Quality Analysis.}}
Our framework generates an average of 2.8 reasoning paths per stock prediction, with 85\% of predictions supported by at least one complete path from the target stock to a textual source. Among the discovered paths, 72\% match at least one rule pattern from the mined rule bank, demonstrating that the exploration successfully leverages learned relational patterns.

The paths exhibit good diversity, with an average of 2.1 distinct relation types per path and 1.8 unique rule patterns per prediction. This variety suggests that the framework captures multiple complementary signals rather than relying on a single reasoning pattern.

\paragraph{\textbf{Evidence Grounding.}}
Examining the temporal relevance of evidence, we find that 65\% of reasoning paths terminate at textual sources published within 7 days of the prediction date, and 86\% within 30 days, ensuring that the supporting evidence is contemporary and particularly relevant for 1-day predictions. The paths connect to diverse information sources including financial news articles, corporate filings, and market reports, providing a broad foundation for predictions.

\subsection{Ablation Studies}
\label{sec:ablation}

We systematically evaluate the contribution of key components through controlled ablation experiments (see Table~\ref{tab:detailed_ablation}).

\paragraph{\textbf{Component Impact Analysis.}}  
Multi-hop reasoning provides the largest performance contribution (2.1\% improvement), confirming that complex financial relationships require multi-step graph traversal. Rule mining contributes 1.1\% improvement, demonstrating that learned relational patterns enhance prediction quality beyond basic graph structure.

Temporal constraints yield 1.4\% improvement, highlighting the importance of respecting information availability dates in financial prediction. Path aggregation provides 0.5\% improvement, showing that combining multiple reasoning paths creates more robust predictions than relying on single paths.

Each component contributes positively to overall performance, validating the design choices in our knowledge graph reasoning framework.

\subsection{Human Expert Analysis}
\label{sec:interpretability_analysis}

\paragraph{\textbf{Path Quality Metrics.}}
We evaluate reasoning path quality using multiple dimensions:
\begin{itemize}
    \item \textbf{Semantic Coherence}: 83\% of paths exhibit logical financial relationships
    \item \textbf{Temporal Consistency}: 91\% respect chronological ordering of events
    \item \textbf{Rule Alignment}: 74\% match at least one mined rule pattern
    \item \textbf{Source Diversity}: Average 2.3 distinct news sources per prediction
\end{itemize}

\paragraph{\textbf{Human Evaluation Protocol.}}
We conducted expert evaluation with financial analyst on 100 randomly selected predictions. Inter-rater reliability (Fleiss' $\kappa = 0.68$) indicates good agreement on path quality. Experts rated 73\% of reasoning paths as ``financially meaningful'' and 64\% as ``actionable for investment decisions.''


\subsection{Comparative Methodology Enhancement}
\label{sec:comparative_analysis}

\paragraph{\textbf{Ensemble Methods Analysis.}}
Combining our knowledge graph reasoning framework with top market data-based baseline (XGBoost) yields 56.1\% accuracy, suggesting complementary signal sources. However, the marginal improvement (0.8$\%$) comes with significant complexity costs, supporting our standalone framework design.

\subsection{Backtesting Analysis}
\label{sec:backtesting}
\paragraph{\textbf{Setup.}}
We evaluate Top-10 buy\&hold baskets (equal-weight, no rebalancing) formed per method over 2023-01-01 to 2024-01-01, alongside an equal-weight all-stocks benchmark. Prices are split and dividend-adjusted and all as-of constraints are enforced.

\begin{figure}[t]
\centering
\includegraphics[width=0.48\textwidth]{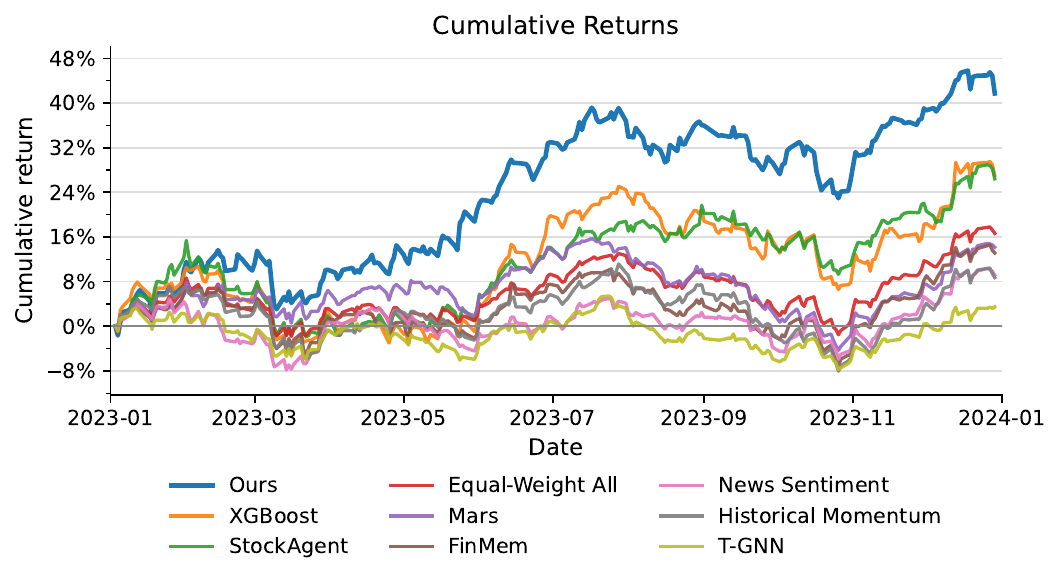}
\caption{Cumulative returns of Top-10 buy\&hold portfolios for all methods (2023-01-01 to 2024-01-01). Our approach dominates over the full horizon and exceeds the equal-weight benchmark.}
\label{fig:backtest_curves}
\end{figure}

\paragraph{\textbf{Results.}}
See Table~\ref{tab:backtest_metrics} and Figure~\ref{fig:backtest_curves}, our method achieves 41.7\% total return with Sharpe 2.00, 18.5\% annualized volatility, \(-11.7\%\) maximum drawdown, and 56\% win rate. The equal-weight benchmark yields 16.7\% total return (Sharpe 1.17). Strong baselines perform lower: XGBoost 27.0\% (1.31), StockAgent 26.4\% (1.43), Mars 14.2\% (0.98), FinMem 13.1\% (0.88), Historical Momentum 8.7\% (0.62), News Sentiment 9.2\% (0.67), and T-GNN 3.5\% (0.33). These results indicate that gains in classification accuracy translate into economically meaningful improvements under identical as-of-date evaluation.

\subsection{Counterfactual Analysis}
\label{sec:counterfactual_refutation}

\paragraph{\textbf{Setup.}}
This analysis is to verify our predicted verdicts rely on the surfaced evidence paths rather than spurious correlations. For each evaluated stock-day, our framework returns a top supporting path
$\mathcal{P}^\star = \langle e_1 \xrightarrow{r_1} \cdots \xrightarrow{r_{k-1}} e_k \rangle$
and a confidence score $s \in [0,1]$ for the verdict $y \in \{\texttt{UP},\texttt{DOWN}\}$.
We perform two counterfactual deletions:
(i) \textbf{Mask-Text}: remove the terminal \texttt{TextSource} node and its incident edge
(\texttt{EXTRACTED\_FROM}), thereby stripping the news evidence tied to the path;
(ii) \textbf{Mask-Edge}: remove the \emph{first critical relation} on the path
(e.g., \texttt{ACQUIRED}, \texttt{PARTNERED}, \texttt{CAUSED\_INCREASE}) that completes the matched rule.
We then recompute the verdict and confidence on the modified graph snapshot. We ensure at most one perturbation per instance, keep search hyperparameters fixed, and recompute predictions on the perturbed snapshot $G_t$. We report Accuracy and F1; values at $r{=}0$ equal the unperturbed evaluation.

\begin{figure}[t]
\centering
\includegraphics[width=0.48\textwidth]{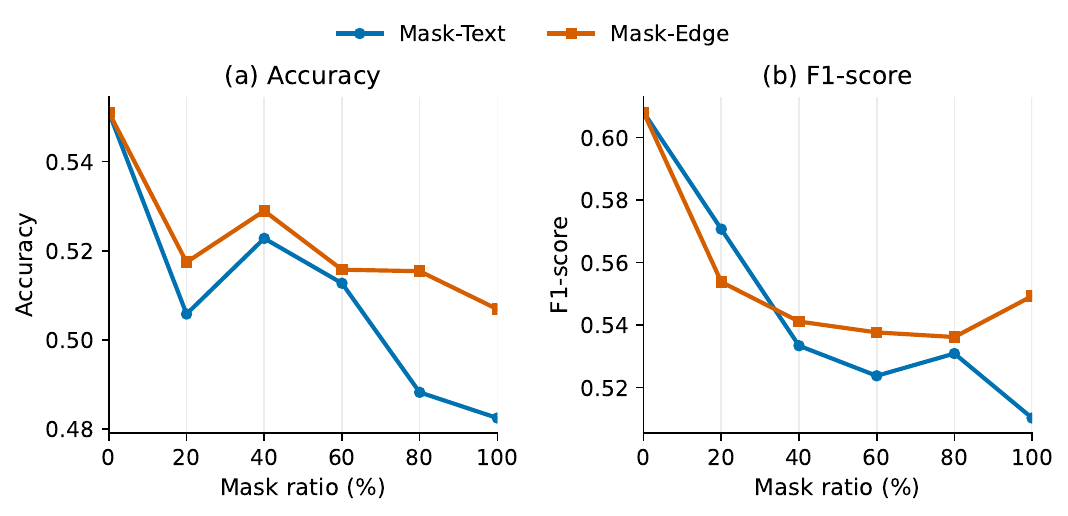}
\caption{Counterfactual Analysis with different deletion ratios for Mask-Tex and Mask-Edge strategies, respectively.}
\label{fig:mask_ratio}
\end{figure}

\paragraph{\textbf{Findings.}}
Figure~\ref{fig:mask_ratio} shows a monotone decline in both metrics with higher mask ratios for both strategies, which validate the effectiveness of evidence paths in our verdicts. \textsc{Mask-Text} degrades faster than \textsc{Mask-Edge}, indicating that news grounding at the terminal node contributes to the decision. Note that at $r{=}100\%$ for \textsc{Mask-Text}, it means that only the news content while no path is used to make the prediction.

\section{Conclusion}

We presented a temporal knowledge-graph reasoning framework named TRACE that unifies rule-guided multi-hop exploration with text-grounded evidence to produce next-day stock movement prediction. On the S\&P~500 task, our method achieves 55.1\% accuracy, 55.7\% precision, 71.5\% recall, and 60.8\% F1, surpassing strong machine-learning, text, graph, and agent-style baselines. In Top-10 buy\&hold backtests over 2023-01-01 to 2024-01-01, it delivers 41.7\% total return with 2.00 Sharpe ratio, outperforming an equal-weight benchmark and competitive baselines. The framework yields path-level explanations while remaining subject to KG coverage and drift; future work will extend to multi-horizon forecasting, calibration, and cross-market validation.

\clearpage
\appendix

\section{Reproducibility Details}
\label{app:repro}
\subsection{Data, Splits, and As-Of Constraints}
\label{app:data}
We study next-day stock movement (close-to-close) for S\&P\,500 constituents in an as-of setting.
Unless otherwise noted, the default split is:
\emph{train} = 2022-01-01\,--\,2023-01-01; \emph{test} = 2023-01-01\,--\,2024-01-01.
For any day $t$, the snapshot $G_t$ and all features include only facts and documents with timestamps $\le t$.
Financial texts (news, filings, transcripts) are deduplicated within a 24h window via ticker-date hashing and text embedding distancing calculation.
Feature normalization is fit on train only and applied to test for market data-based baselines.

\subsection{Model and Search Settings}
\label{app:settings}
We run a rule-guided search over the as-of snapshot $G_t$ with beam width $K{=}8$, max depth $D_{\max}{=}3$,
hypothesis threshold $\tau_{\mathrm{hyp}}{=}0.60$, and at most 10 scored paths per instance (top-$M$ evidence budget $M{=}5$ by default).
Textual relevance is computed with Qwen3-Embedding-8B for dense representations of \texttt{TextSource} nodes and query snippets.

We employ Qwen QwQ-32B as a frozen controller in two places:
\begin{enumerate}
  \item \textsc{LLMRelationSelector}: given a set of candidate one-hop extensions from the current frontier, the LLM filters expansions that are (i) semantically compatible with the news snippet and (ii) consistent with the partially matched rule prefix. LLMs reduce hub and semantic drift effectively.
  \item \textsc{LLMValidate}: for each completed (path, rule) pair $(P,\rho)$, the LLM verifies label consistency with the news content and yields $(\hat{\ell}_P, p_P)$, where $p_P\in[0,1]$ is a calibrated plausibility score used in evidence fusion.
\end{enumerate}

\subsection{Evaluation and Significance}
\label{app:eval}
This section defines the \emph{classification} and \emph{backtesting} metrics used in the paper. For each (ticker, day) instance $i\in\{1,\dots,N\}$ let $y_i\in\{\texttt{UP},\texttt{DOWN}\}$ be the ground-truth next-day label (close-to-close), and $\hat{y}_i$ the model's prediction.

\paragraph{Classification.}
We report the standard point metrics on the pooled (ticker, day) set:
\[
\mathrm{Accuracy}=\frac{\mathrm{TP}+\mathrm{TN}}{N},\qquad
\mathrm{Precision}=\frac{\mathrm{TP}}{\mathrm{TP}+\mathrm{FP}},
\]
\[
\mathrm{Recall}=\frac{\mathrm{TP}}{\mathrm{TP}+\mathrm{FN}},\qquad
\mathrm{F1}=2\cdot \frac{\mathrm{Precision}\cdot\mathrm{Recall}}{\mathrm{Precision}+\mathrm{Recall}}\;.
\]

\paragraph{Backtesting.}
Let $\mathcal{B}$ be the Top-10, equal-weight, long-only basket chosen at the \emph{start} of the test window. For each trading day $t=1,\dots,T$ let $r_{j,t}$ be the simple return of asset $j\in\mathcal{B}$ from close$_{t-1}$ to close$_t$:
\[
r_{j,t}=\frac{P_{j,t}-P_{j,t-1}}{P_{j,t-1}}\;.
\]
Daily portfolio return (equal weights $w_j=1/|\mathcal{B}|$):
\[
r_t=\sum_{j\in\mathcal{B}} w_j\, r_{j,t}\quad\text{with }w_j=\tfrac{1}{10}.
\]
Denote the daily risk-free rate by $r^{(f)}_t$ (set to $0$ unless stated).

\paragraph{Total Return (cumulative).}
\[
\mathrm{TotRet}=\prod_{t=1}^{T}(1+r_t)-1.
\]

\paragraph{Annualized Volatility.}
Let $\bar{r}=\frac{1}{T}\sum_{t=1}^{T} r_t$ and $s=\sqrt{\frac{1}{T-1}\sum_{t=1}^{T}(r_t-\bar{r})^2}$ be the sample std.\ of daily returns. Then
\[
\mathrm{AnnVol}= s \cdot \sqrt{252}.
\]

\paragraph{Sharpe Ratio (annualized).}
Using daily \emph{excess} returns $\tilde{r}_t=r_t-r^{(f)}_t$, with sample mean $\bar{\tilde{r}}$ and std.\ $s_{\tilde{r}}$:
\[
\mathrm{Sharpe}= \frac{\bar{\tilde{r}}\cdot 252}{s_{\tilde{r}}\cdot \sqrt{252}}
= \sqrt{252}\cdot \frac{\bar{\tilde{r}}}{s_{\tilde{r}}}.
\]

\paragraph{Maximum Drawdown.}
Let the cumulative equity curve be $C_t=\prod_{u=1}^{t}(1+r_u)$ and running peak $H_t=\max_{u\le t} C_u$. Drawdown at $t$ is $DD_t=\frac{C_t}{H_t}-1$. The reported statistic is
\[
\mathrm{MaxDD}=\min_{t\in\{1,\dots,T\}} DD_t\quad(\text{a non-positive number}).
\]

\paragraph{Win Rate.}
\[
\mathrm{WinRate}=\frac{1}{T}\sum_{t=1}^{T}\mathbb{I}[r_t>0]\;.
\]

At the test window start, each method forms an equal-weight, long-only basket of its Top-10 tickers and holds (no rebalancing). Unless specified, dividends, fees, and slippage are excluded.

\subsection{Counterfactual and Mask-Ratio Studies}
\label{app:theweb_cf}
\paragraph{Counterfactual deletions.}
(\textsc{Mask-Text}) remove the terminal \texttt{TextSource}, which denotes news and event content, and its \texttt{EXTRACTED\_FROM} edge on the top path; 
(\textsc{Mask-Edge}) remove the first rule-completing relation (e.g., \texttt{ACQUIRED}, \texttt{PARTNERED}, \texttt{CAUSED\_INCREASE}).
We then recompute verdicts with confidences on the perturbed $G_t$ and report (i) Accuracy and (ii) F1-Score.
\paragraph{Mask-ratio sweeps.}
For $r \in \{0,20,40,60,80,100\}\%$, perturb a random $r\%$ of instances (one perturbation per instance) and re-evaluate.

\section{Ethical and Use Considerations}
\label{app:theweb_ethics}
We operate on public, time-stamped information and enforce strict as-of filtering to avoid look-ahead bias.
Models are \emph{assistive} and not financial advice. Interpretability is provided via rule and news paths. Backtesting is illustrative and sensitive to costs and slippage.

\section{Algorithm Pipeline}
We illustrate the end-to-end reasoning pipeline that underlies all results in  Algorithm~\ref{alg:rule_guided_tkg} for more technique details.

\begin{algorithm}[t]
\caption{Rule-Guided Temporal KG Reasoning}
\label{alg:rule_guided_tkg}
\begin{algorithmic}[1]
\Require News $x$ with time $\tau(x)$; KG $G_t=(E_t,R_t,F_t)$ with $t\!\ge\!\tau(x)$; rules $R^\ast$; beam $K$; max depth $D_{\max}$; threshold $\tau_{\mathrm{hyp}}$; evidence budget $M$; fusion weight $\alpha\in[0,1]$
\Ensure $y\in\{\texttt{UP},\texttt{DOWN}\}$ with confidence $\mathrm{conf}$ and evidence $\mathcal{P}$

\State $S\!\gets\!\textsc{ExtractEntitiesAndRelations}(x)$;\quad $G_t\!\gets\!\textsc{GroundToKG}(S,G_t)$
\State $\textsc{Score}(P)\!=\!\big\langle \mathbb{I}[\textsc{CompletesHighConfRule}(P,R^\ast,\tau_{\mathrm{hyp}})],$\\
$~\textsc{BestRuleCoverage}(P,R^\ast),~\textsc{Recency}(P),~\textsc{AntiHubPenalty}(P),~-|P|\big\rangle$

\For{$d\!=\!1$ \textbf{to} $D_{\max}$}
  \State $\mathsf{Cands}\!\gets\!\emptyset$
  \ForAll{$P\in\mathsf{Beam}$}
    \ForAll{neighbors $(\textsc{Head}(P)\xrightarrow{r}v)$ valid at $t$}
      \If{$\textsc{IsPrefixOfSomeRule}(P\!\cdot\!(r,v),R^\ast)$}
        \State $P'\!\gets\!P\!\cdot\!(r,v)$; add $P'$ to $\mathsf{Cands}$
        \If{$\textsc{CompletesHighConfRule}(P',R^\ast,\tau_{\mathrm{hyp}})$}
          \State $(\rho,\ell,\mathrm{conf}(\rho))\!\gets\!\textsc{MatchedRule}(P',R^\ast)$; add $(P',\rho,\mathrm{conf}(\rho),\ell)$ to $\mathcal{H}$
        \EndIf
      \EndIf
    \EndFor
  \EndFor
  \If{$\mathsf{Cands}\!=\!\emptyset$} \textbf{break} \EndIf
  \State $\mathsf{Cands}\!\gets\!\textsc{LLMRelationSelector}(\mathsf{Cands})$;\quad $\mathsf{Beam}\!\gets\!\textsc{TopK}(\mathsf{Cands},K,\text{lexi }\textsc{Score})$
\EndFor

\If{$\mathcal{H}\!=\!\emptyset$} \State \Return $(\texttt{DOWN},0,\emptyset)$ \EndIf
\State $\mathrm{confUP}\!\gets\!\max_{(P,\rho,\cdot,\ell)\in\mathcal{H},\,\ell=\texttt{UP}}\mathrm{conf}(\rho)$;\quad
       $\mathrm{confDOWN}\!\gets\!\max_{(P,\rho,\cdot,\ell)\in\mathcal{H},\,\ell=\texttt{DOWN}}\mathrm{conf}(\rho)$
\If{$\mathrm{confUP}\!\ge\!\mathrm{confDOWN}$} \State $y\!\gets\!\texttt{UP}$;\; $\mathrm{conf}\!\gets\!\mathrm{confUP}$
\Else \State $y\!\gets\!\texttt{DOWN}$;\; $\mathrm{conf}\!\gets\!\mathrm{confDOWN}$ \EndIf

\State $\mathcal{E}\!\gets\!\emptyset$;\; \ForAll{$(P,\rho,\mathrm{conf}(\rho),\ell)\in\mathcal{H}$}
  \State $(\hat{\ell}_P,p_P)\!\gets\!\textsc{LLMValidate}(x,P,\rho)$; \textbf{if} $\hat{\ell}_P\!=\!\ell$ \textbf{then} add $(P,\rho,\mathrm{conf}(\rho),\ell,p_P)$ to $\mathcal{E}$
\EndFor
\State $\mathcal{E}_y\!\gets\!\{e\in\mathcal{E}:\ell=y\}$;\quad $\mathcal{P}\!\gets\!\textsc{TopMBy}\big(\mathcal{E}_y,M,~\alpha\,\mathrm{conf}(\rho)+(1-\alpha)\,p_P\big)$
\State \Return $(y,\mathrm{conf},\mathcal{P})$
\end{algorithmic}
\end{algorithm}

\clearpage
\bibliographystyle{ACM-Reference-Format}
\bibliography{references}

@String{Computing = "Computing" }

@String{Computer = "{IEEE} Computer" }

@String{Springer = "Springer-Verlag" }

@BOOK{test,
   author = "Donald E. Knuth",
   title = "Seminumerical Algorithms",
   volume = 2,
   series = "The Art of Computer Programming",
   publisher = "Addison-Wesley",
   address = "Reading, MA",
   edition = "2nd",
   month = "10~" # jan,
   year = "1981",
}

@ArtifactSoftware{R,
    title = {R: A Language and Environment for Statistical Computing},
    author = {{R Core Team}},
    organization = {R Foundation for Statistical Computing},
    address = {Vienna, Austria},
    year = {2019},
    url = {https://www.R-project.org/},
}

@inproceedings{trivedi2017know,
  title={Know-evolve: Deep temporal reasoning for dynamic knowledge graphs},
  author={Trivedi, Rakshit and Dai, Hanjun and Wang, Yichen and Song, Le},
  booktitle={international conference on machine learning},
  pages={3462--3471},
  year={2017},
  organization={PMLR}
}

@article{jin2019recurrent,
  title={Recurrent event network: Autoregressive structure inference over temporal knowledge graphs},
  author={Jin, Woojeong and Qu, Meng and Jin, Xisen and Ren, Xiang},
  journal={arXiv preprint arXiv:1904.05530},
  year={2019}
}

@inproceedings{dasgupta2018hyte,
  title={Hyte: Hyperplane-based temporally aware knowledge graph embedding},
  author={Dasgupta, Shib Sankar and Ray, Swayambhu Nath and Talukdar, Partha},
  booktitle={Proceedings of the 2018 conference on empirical methods in natural language processing},
  pages={2001--2011},
  year={2018}
}

@article{kim2019hats,
  title={Hats: A hierarchical graph attention network for stock movement prediction},
  author={Kim, Raehyun and So, Chan Ho and Jeong, Minbyul and Lee, Sanghoon and Kim, Jinkyu and Kang, Jaewoo},
  journal={arXiv preprint arXiv:1908.07999},
  year={2019}
}

@article{feng2019temporal,
  title={Temporal relational ranking for stock prediction},
  author={Feng, Fuli and He, Xiangnan and Wang, Xiang and Luo, Cheng and Liu, Yiqun and Chua, Tat-Seng},
  journal={ACM Transactions on Information Systems (TOIS)},
  volume={37},
  number={2},
  pages={1--30},
  year={2019},
  publisher={ACM New York, NY, USA}
}

@inproceedings{qian2024mdgnn,
  title={Mdgnn: Multi-relational dynamic graph neural network for comprehensive and dynamic stock investment prediction},
  author={Qian, Hao and Zhou, Hongting and Zhao, Qian and Chen, Hao and Yao, Hongxiang and Wang, Jingwei and Liu, Ziqi and Yu, Fei and Zhang, Zhiqiang and Zhou, Jun},
  booktitle={Proceedings of the AAAI Conference on Artificial Intelligence},
  volume={38},
  number={13},
  pages={14642--14650},
  year={2024}
}

@article{liu2024risk,
  title={Risk identification and management through knowledge Association: A financial event evolution knowledge graph approach},
  author={Liu, Zhenghao and Zhang, Zhijian and Zeng, Xi},
  journal={Expert Systems with Applications},
  volume={252},
  pages={123999},
  year={2024},
  publisher={Elsevier}
}

@inproceedings{li2024findkg,
  title={Findkg: Dynamic knowledge graphs with large language models for detecting global trends in financial markets},
  author={Li, Xiaohui Victor and Sanna Passino, Francesco},
  booktitle={Proceedings of the 5th ACM international conference on AI in finance},
  pages={573--581},
  year={2024}
}

@article{wang2023fingpt,
  title={Fingpt: Instruction tuning benchmark for open-source large language models in financial datasets},
  author={Wang, Neng and Yang, Hongyang and Wang, Christina Dan},
  journal={arXiv preprint arXiv:2310.04793},
  year={2023}
}

@article{wu2023bloomberggpt,
  title={Bloomberggpt: A large language model for finance},
  author={Wu, Shijie and Irsoy, Ozan and Lu, Steven and Dabravolski, Vadim and Dredze, Mark and Gehrmann, Sebastian and Kambadur, Prabhanjan and Rosenberg, David and Mann, Gideon},
  journal={arXiv preprint arXiv:2303.17564},
  year={2023}
}

@article{galarraga2015fast,
  title={Fast rule mining in ontological knowledge bases with AMIE $$+ $$},
  author={Gal{\'a}rraga, Luis and Teflioudi, Christina and Hose, Katja and Suchanek, Fabian M},
  journal={The VLDB Journal},
  volume={24},
  number={6},
  pages={707--730},
  year={2015},
  publisher={Springer}
}

@article{meilicke2020reinforced,
  title={Reinforced anytime bottom up rule learning for knowledge graph completion},
  author={Meilicke, Christian and Chekol, Melisachew Wudage and Fink, Manuel and Stuckenschmidt, Heiner},
  journal={arXiv preprint arXiv:2004.04412},
  year={2020}
}

@article{yang2017differentiable,
  title={Differentiable learning of logical rules for knowledge base reasoning},
  author={Yang, Fan and Yang, Zhilin and Cohen, William W},
  journal={Advances in neural information processing systems},
  volume={30},
  year={2017}
}

@article{qu2020rnnlogic,
  title={Rnnlogic: Learning logic rules for reasoning on knowledge graphs},
  author={Qu, Meng and Chen, Junkun and Xhonneux, Louis-Pascal and Bengio, Yoshua and Tang, Jian},
  journal={arXiv preprint arXiv:2010.04029},
  year={2020}
}

@article{saeed2021rulebert,
  title={RuleBERT: Teaching soft rules to pre-trained language models},
  author={Saeed, Mohammed and Ahmadi, Naser and Nakov, Preslav and Papotti, Paolo},
  journal={arXiv preprint arXiv:2109.13006},
  year={2021}
}

@article{sun2023think,
  title={Think-on-graph: Deep and responsible reasoning of large language model on knowledge graph},
  author={Sun, Jiashuo and Xu, Chengjin and Tang, Lumingyuan and Wang, Saizhuo and Lin, Chen and Gong, Yeyun and Ni, Lionel M and Shum, Heung-Yeung and Guo, Jian},
  journal={arXiv preprint arXiv:2307.07697},
  year={2023}
}

@article{ma2024think,
  title={Think-on-graph 2.0: Deep and faithful large language model reasoning with knowledge-guided retrieval augmented generation},
  author={Ma, Shengjie and Xu, Chengjin and Jiang, Xuhui and Li, Muzhi and Qu, Huaren and Yang, Cehao and Mao, Jiaxin and Guo, Jian},
  journal={arXiv preprint arXiv:2407.10805},
  year={2024}
}

@article{luo2024graph,
  title={Graph-constrained reasoning: Faithful reasoning on knowledge graphs with large language models},
  author={Luo, Linhao and Zhao, Zicheng and Haffari, Gholamreza and Li, Yuan-Fang and Gong, Chen and Pan, Shirui},
  journal={arXiv preprint arXiv:2410.13080},
  year={2024}
}

@article{zhang2024chain,
  title={Chain-of-knowledge: Integrating knowledge reasoning into large language models by learning from knowledge graphs},
  author={Zhang, Yifei and Wang, Xintao and Liang, Jiaqing and Xia, Sirui and Chen, Lida and Xiao, Yanghua},
  journal={arXiv preprint arXiv:2407.00653},
  year={2024}
}

@article{liao2023gentkg,
  title={Gentkg: Generative forecasting on temporal knowledge graph with large language models},
  author={Liao, Ruotong and Jia, Xu and Li, Yangzhe and Ma, Yunpu and Tresp, Volker},
  journal={arXiv preprint arXiv:2310.07793},
  year={2023}
}

@article{yu2025rag,
  title={Rag-kg-il: A multi-agent hybrid framework for reducing hallucinations and enhancing llm reasoning through rag and incremental knowledge graph learning integration},
  author={Yu, Hong Qing and McQuade, Frank},
  journal={arXiv preprint arXiv:2503.13514},
  year={2025}
}

@article{yang2020finbert,
  title={Finbert: A pretrained language model for financial communications},
  author={Yang, Yi and Uy, Mark Christopher Siy and Huang, Allen},
  journal={arXiv preprint arXiv:2006.08097},
  year={2020}
}

@article{araci2019finbert,
  title={Finbert: Financial sentiment analysis with pre-trained language models},
  author={Araci, Dogu},
  journal={arXiv preprint arXiv:1908.10063},
  year={2019}
}

@inproceedings{ding2015deep,
  title={Deep learning for event-driven stock prediction.},
  author={Ding, Xiao and Zhang, Yue and Liu, Ting and Duan, Junwen},
  booktitle={Ijcai},
  volume={15},
  pages={2327--2333},
  year={2015}
}

@inproceedings{zhou2021informer,
  title={Informer: Beyond efficient transformer for long sequence time-series forecasting},
  author={Zhou, Haoyi and Zhang, Shanghang and Peng, Jieqi and Zhang, Shuai and Li, Jianxin and Xiong, Hui and Zhang, Wancai},
  booktitle={Proceedings of the AAAI conference on artificial intelligence},
  volume={35},
  number={12},
  pages={11106--11115},
  year={2021}
}

@article{bai2018empirical,
  title={An empirical evaluation of generic convolutional and recurrent networks for sequence modeling},
  author={Bai, Shaojie and Kolter, J Zico and Koltun, Vladlen},
  journal={arXiv preprint arXiv:1803.01271},
  year={2018}
}

@inproceedings{ariyo2014stock,
  title={Stock price prediction using the ARIMA model},
  author={Ariyo, Adebiyi A and Adewumi, Adewumi O and Ayo, Charles K},
  booktitle={2014 UKSim-AMSS 16th international conference on computer modelling and simulation},
  pages={106--112},
  year={2014},
  organization={IEEE}
}

@article{fischer2018deep,
  title={Deep learning with long short-term memory networks for financial market predictions},
  author={Fischer, Thomas and Krauss, Christopher},
  journal={European journal of operational research},
  volume={270},
  number={2},
  pages={654--669},
  year={2018},
  publisher={Elsevier}
}

@inproceedings{nelson2017stock,
  title={Stock market's price movement prediction with LSTM neural networks},
  author={Nelson, David MQ and Pereira, Adriano CM and De Oliveira, Renato A},
  booktitle={2017 International joint conference on neural networks (IJCNN)},
  pages={1419--1426},
  year={2017},
  organization={Ieee}
}

@inproceedings{yoo2021accurate,
  title={Accurate multivariate stock movement prediction via data-axis transformer with multi-level contexts},
  author={Yoo, Jaemin and Soun, Yejun and Park, Yong-chan and Kang, U},
  booktitle={Proceedings of the 27th ACM SIGKDD Conference on Knowledge Discovery \& Data Mining},
  pages={2037--2045},
  year={2021}
}

@article{long2019deep,
  title={Deep learning-based feature engineering for stock price movement prediction},
  author={Long, Wen and Lu, Zhichen and Cui, Lingxiao},
  journal={Knowledge-Based Systems},
  volume={164},
  pages={163--173},
  year={2019},
  publisher={Elsevier}
}

@article{feng2018enhancing,
  title={Enhancing stock movement prediction with adversarial training},
  author={Feng, Fuli and Chen, Huimin and He, Xiangnan and Ding, Ji and Sun, Maosong and Chua, Tat-Seng},
  journal={arXiv preprint arXiv:1810.09936},
  year={2018}
}

@inproceedings{xu2018stock,
  title={Stock movement prediction from tweets and historical prices},
  author={Xu, Yumo and Cohen, Shay B},
  booktitle={Proceedings of the 56th Annual Meeting of the Association for Computational Linguistics (Volume 1: Long Papers)},
  pages={1970--1979},
  year={2018}
}

@article{nguyen2015sentiment,
  title={Sentiment analysis on social media for stock movement prediction},
  author={Nguyen, Thien Hai and Shirai, Kiyoaki and Velcin, Julien},
  journal={Expert Systems with Applications},
  volume={42},
  number={24},
  pages={9603--9611},
  year={2015},
  publisher={Elsevier}
}

@inproceedings{ding2020hierarchical,
  title={Hierarchical multi-scale Gaussian transformer for stock movement prediction.},
  author={Ding, Qianggang and Wu, Sifan and Sun, Hao and Guo, Jiadong and Guo, Jian},
  booktitle={Ijcai},
  pages={4640--4646},
  year={2020}
}

@article{ding2024tradexpert,
  title={Tradexpert: Revolutionizing trading with mixture of expert llms},
  author={Ding, Qianggang and Shi, Haochen and Guo, Jiadong and Liu, Bang},
  journal={arXiv preprint arXiv:2411.00782},
  year={2024}
}

@inproceedings{huang2019knowledge,
  title={Knowledge graph embedding based question answering},
  author={Huang, Xiao and Zhang, Jingyuan and Li, Dingcheng and Li, Ping},
  booktitle={Proceedings of the twelfth ACM international conference on web search and data mining},
  pages={105--113},
  year={2019}
}

@article{rossi2021knowledge,
  title={Knowledge graph embedding for link prediction: A comparative analysis},
  author={Rossi, Andrea and Barbosa, Denilson and Firmani, Donatella and Matinata, Antonio and Merialdo, Paolo},
  journal={ACM Transactions on Knowledge Discovery from Data (TKDD)},
  volume={15},
  number={2},
  pages={1--49},
  year={2021},
  publisher={ACM New York, NY, USA}
}

@article{lin2018multi,
  title={Multi-hop knowledge graph reasoning with reward shaping},
  author={Lin, Xi Victoria and Socher, Richard and Xiong, Caiming},
  journal={arXiv preprint arXiv:1808.10568},
  year={2018}
}

@article{chen2025mars,
  title={MARS: A Meta-Adaptive Reinforcement Learning Framework for Risk-Aware Multi-Agent Portfolio Management},
  author={Chen, Jiayi and Li, Jing and Wang, Guiling},
  journal={arXiv preprint arXiv:2508.01173},
  year={2025}
}

@article{zhang2024ai,
  title={When ai meets finance (stockagent): Large language model-based stock trading in simulated real-world environments},
  author={Zhang, Chong and Liu, Xinyi and Zhang, Zhongmou and Jin, Mingyu and Li, Lingyao and Wang, Zhenting and Hua, Wenyue and Shu, Dong and Zhu, Suiyuan and Jin, Xiaobo and others},
  journal={arXiv preprint arXiv:2407.18957},
  year={2024}
}

@article{yu2025finmem,
  title={Finmem: A performance-enhanced llm trading agent with layered memory and character design},
  author={Yu, Yangyang and Li, Haohang and Chen, Zhi and Jiang, Yuechen and Li, Yang and Suchow, Jordan W and Zhang, Denghui and Khashanah, Khaldoun},
  journal={IEEE Transactions on Big Data},
  year={2025},
  publisher={IEEE}
}

@article{vuong2022stock,
  title={Stock-price forecasting based on XGBoost and LSTM.},
  author={Vuong, Pham Hoang and Dat, Trinh Tan and Mai, Tieu Khoi and Uyen, Pham Hoang and others},
  journal={Computer Systems Science \& Engineering},
  volume={40},
  number={1},
  year={2022}
}

@article{yang2025qwen3,
  title={Qwen3 technical report},
  author={Yang, An and Li, Anfeng and Yang, Baosong and Zhang, Beichen and Hui, Binyuan and Zheng, Bo and Yu, Bowen and Gao, Chang and Huang, Chengen and Lv, Chenxu and others},
  journal={arXiv preprint arXiv:2505.09388},
  year={2025}
}

@inproceedings{xiang2022temporal,
  title={Temporal and heterogeneous graph neural network for financial time series prediction},
  author={Xiang, Sheng and Cheng, Dawei and Shang, Chencheng and Zhang, Ying and Liang, Yuqi},
  booktitle={Proceedings of the 31st ACM international conference on information \& knowledge management},
  pages={3584--3593},
  year={2022}
}

@BOOK{Jurafsky2009-ei,
  title     = "Speech and Language Processing : An Introduction to Natural
               Language Processing, Computational Linguistics, and Speech
               Recognition. 2",
  author    = "Jurafsky, Daniel and Martin, James H",
  publisher = "Pearson Education International",
  year      =  2009
}

@inproceedings{10.1145/3539618.3591698,
author = {Atif, Farah and El Khatib, Ola and Difallah, Djellel},
title = {BeamQA: Multi-hop Knowledge Graph Question Answering with Sequence-to-Sequence Prediction and Beam Search},
year = {2023},
isbn = {9781450394086},
publisher = {Association for Computing Machinery},
address = {New York, NY, USA},
url = {https://doi.org/10.1145/3539618.3591698},
doi = {10.1145/3539618.3591698},
booktitle = {Proceedings of the 46th International ACM SIGIR Conference on Research and Development in Information Retrieval},
pages = {781–790},
numpages = {10},
location = {Taipei, Taiwan},
series = {SIGIR '23}
}

@article{sun2023beamsearchqa,
  title={Beamsearchqa: Large language models are strong zero-shot qa solver},
  author={Sun, Hao and Liu, Xiao and Gong, Yeyun and Dong, Anlei and Lu, Jingwen and Zhang, Yan and Jiang, Daxin and Yang, Linjun and Majumder, Rangan and Duan, Nan},
  journal={arXiv preprint arXiv:2305.14766},
  year={2023}
}

@inproceedings{liumakes,
  title={What Makes Good Data for Alignment? A Comprehensive Study of Automatic Data Selection in Instruction Tuning},
  author={Liu, Wei and Zeng, Weihao and He, Keqing and Jiang, Yong and He, Junxian},
  booktitle={The Twelfth International Conference on Learning Representations}
}

@article{jiang2023structgpt,
  title={Structgpt: A general framework for large language model to reason over structured data},
  author={Jiang, Jinhao and Zhou, Kun and Dong, Zican and Ye, Keming and Zhao, Wayne Xin and Wen, Ji-Rong},
  journal={arXiv preprint arXiv:2305.09645},
  year={2023}
}

@article{wang2023boosting,
  title={Boosting language models reasoning with chain-of-knowledge prompting},
  author={Wang, Jianing and Sun, Qiushi and Li, Xiang and Gao, Ming},
  journal={arXiv preprint arXiv:2306.06427},
  year={2023}
}

@article{baek2023knowledge,
  title={Knowledge-augmented language model prompting for zero-shot knowledge graph question answering},
  author={Baek, Jinheon and Aji, Alham Fikri and Saffari, Amir},
  journal={arXiv preprint arXiv:2306.04136},
  year={2023}
}

@article{xie2022unifiedskg,
  title={Unifiedskg: Unifying and multi-tasking structured knowledge grounding with text-to-text language models},
  author={Xie, Tianbao and Wu, Chen Henry and Shi, Peng and Zhong, Ruiqi and Scholak, Torsten and Yasunaga, Michihiro and Wu, Chien-Sheng and Zhong, Ming and Yin, Pengcheng and Wang, Sida I and others},
  journal={arXiv preprint arXiv:2201.05966},
  year={2022}
}
\end{document}